# Enhancing silicon solar cells with singlet fission: the case for Förster resonant energy transfer using a quantum dot intermediate

Stefan Wil Tabernig[+], Benjamin Daiber[+], Tianyi Wang, Bruno Ehrler*

+ … contributed equally

Center for Nanophotonics, AMOLF, Science Park 104, 1098 XG Amsterdam, The Netherlands

*E-mail: b.ehrler@amolf.nl

Abstract

One way for solar cell efficiencies to overcome the Shockley-Queisser limit is downconversion of high-energy photons using singlet fission (SF) in polyacenes like tetracene (Tc). SF enables generation of multiple excitons from the high-energy photons which can be harvested in combination with Si. In this work we investigate the use of lead sulfide quantum dots (PbS QDs) with a band gap close to Si as an interlayer that allows Förster Resonant Energy Transfer (FRET) from Tc to Si, a process that would be spin-forbidden without the intermediate QD step. We investigate how the conventional FRET model, most commonly applied to the description of molecular interactions, can be modified to describe the geometry of QDs between Tc and Si and how the distance between QD and Si, and the QD bandgap affects the FRET efficiency. By extending the acceptor dipole in the FRET model to a 2D plane, and to the bulk, we see a relaxation of the distance dependence of transfer. Our results indicate that FRET efficiencies from PbS QDs to Si well above 50 % are be possible at very short, but possibly realistic distances of around 1 nm, even for quantum dots with relatively low photoluminescence quantum yield.

## Introduction

The domination of the solar cell market by silicon led to the search of add-ons that could increase efficiency while also maintaining low cost. One possible way to increase efficiency is by downconverting high-energy light using an organic material that exhibits singlet fission.

In a single-junction solar cell, photons with energy above the bandgap can excite an electron into the conduction band. Excess energy is lost, as the charge carriers quickly thermalize to the band edge. Downconversion schemes minimize the energy lost by thermalization, by converting high-energy photons to lower-energy charge carriers. Downconversion via singlet fission can improve on the single-junction Shockley-Queisser[1,2] efficiency limit, raising it from 33.7 % to 44.4 %[3].

Singlet fission in organic semiconductors describes the conversion of a singlet exciton into two triplet excitons, conserving spin. In tetracene, singlet fission is faster (90 ps)[4] than other decay channels which leads to a yield of almost two triplet excitons per absorbed photon. The resulting triplet excitons cannot relax radiatively to the singlet ground state, as this process is spin-forbidden, leading to a long triplet lifetime. In tetracene, the energy of the triplet excitons ($1.25\ eV$)[5] is close to the bandgap of silicon ($1.12\ eV$), allowing in principle for the triplet excitons to be injected into silicon (Si). In one possible realization, the triplet exciton energy is first transferred into a lead sulphide (PbS) quantum dot (QD)[6] interlayer and subsequently transferred into Si[7,8] (see fig.1Figure 1). Once the triplet exciton is in the QD, the presence of lead with strong spin-orbit coupling leads to intersystem crossing of singlet and triplet states. The spin triplet and singlet excitons are energy degenerate ($\approx 3\ meV$)[9] which makes the spin mixing efficient. Hence, the exciton can decay radiatively, in principle allowing for energy transfer into Si via photon emission, or Förster resonant energy transfer (FRET). Transfer into lead sulfide[6] and lead selenide[10] QDs was recently demonstrated with high efficiency (> 90%)[6]. While energy transfer from core/shell CdSe/ZnS QDs into c-Si[7] as well as inter-QD FRET for cases where energy was transferred amongst the same QD species[11–13] and between two different QD species[12,14] has been demonstrated, energy transfer from PbS QDs into Si with a QD bandgap close to the one of c-Si remains to be shown.

One of the processes competing with FRET is the emission of photons by the QDs and the re-absorption in Si. For that process to be efficient, careful light management to funnel photons into silicon is required. In addition, the low external quantum efficiency (EQE) of the Si cell near the indirect band edge might somewhat limit the achievable efficiency. Direct energy transfer in the form of FRET would be an elegant solution to allow for higher efficiency, as FRET can outcompete radiative energy transfer at distances smaller than the system-specific Förster distance $R_0$, which is around $8\,nm$ in the case of FRET between PbS QDs[13,15]. Once the exciton resides in Si it will contribute to charge generation, as the extraction efficiency of state-of-the-art Si solar cells is close to unity. Thus, the singlet fission-FRET geometry could lead to additional current in Si solar cells, if short distances between the donor and acceptor can be achieved.

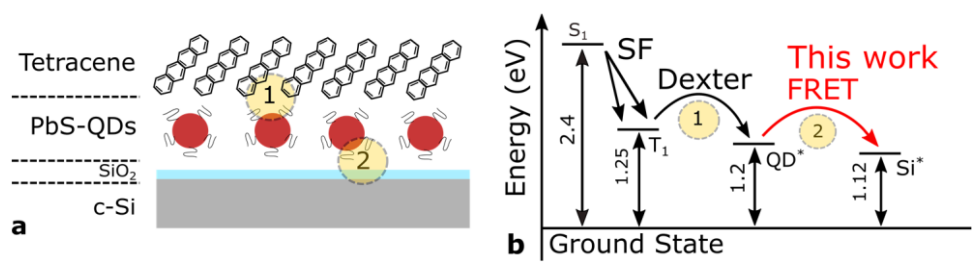

*Figure 1: **(a)** Illustration of the singlet fission-FRET geometry. A Tc-layer lies on top of the PbS-QD (+ligands) monolayer, which is on top of c-Si. The two yellow circles indicate the two energy transfer steps, namely Tc→QD (1) and QD→Si (2). **(b)** The Jablonski diagram, with the FRET process between QDs and Si highlighted in red. $S_1$ and $T_1$ correspond to the first excited singlet and triplet state in tetracene, respectively. The excited states of the QD and Si are indicated by $QD^*$ and $Si^*$.*

Here we establish the theoretical requirements for FRET between PbS QDs and Si, considering the QD bandgap, the distance between Si and QDs, and the geometry of the interface. We find that FRET can be efficient when the QDs are within 1.5 nm to the surface of Si, even for QDs with a bandgap close to the Si bandgap. This is a much shorter distance compared to inter-QD FRET or organic molecules, mostly because the molar absorption coefficient of Si is very low near the band edge. We further find that the distance dependence is somewhat relaxed when considering the Si surface as a plane or bulk acceptor. Finally, we lay out the path to prepare the Si surface to allow for efficient FRET from tetracene into Si. Once efficient transfer of energy between QDs and Si can be achieved experimentally, singlet fission could provide a direct path towards more efficient Si solar cells with minimal need for changes of the Si cell geometry.

## FRET

The FRET efficiency of excitons from QDs into Si, $\eta_{FRET}$, is defined in eq.5. The main goal of this work is to determine how $\eta_{FRET}$ depends on donor acceptor distance, on the bandgap of the QDs, and on the geometry of the system. The FRET efficiency $\eta_{FRET}$ compares the FRET rate $k_{FRET}$ to all the competing rates, defined as the exciton decay rate of the QD donor in absence of the silicon acceptor $k_{D,0}$[16].

$$\eta_{FRET} = \frac{k_{FRET}}{k_{D,0} + k_{FRET}}$$

Where $k_{D,0} = 1/\tau_{D,0}$ and $\tau_{D,0}$ represents the donor exciton lifetime in absence of an acceptor.

FRET is a distance-dependent energy transfer mechanism between two molecules, which are approximated to be point dipoles. Förster derived an expression for the FRET rate[17] which depends on the emission spectrum of the donor, absorption spectrum of the acceptor, donor lifetime, and donor acceptor distance. The classical as well as quantum mechanical approach both lead to eq.2[16,17]:

$$k_{FRET}(R_{DA}) = \frac{1}{\tau_{D,0}} \left(\frac{R_0}{R_{DA}}\right)^6$$

Here, $R_{DA}$ represents the distance between donor and acceptor and $R_0$ is the Förster distance. $R_0$ determines how strongly the FRET rate depends on the distance and is given by eq.3 [16]:

$$R_0^6 = \frac{9000 \ln(10)}{128\,\pi^5\,N_A} * \frac{Q_D \kappa^2 J}{n^4}$$

In eq.3, the prefactor summarizes several numerical constants and the Avogadro's number $N_A$. $Q_D$ is the donor photoluminescence quantum yield (PLQY), $\kappa^2$ is a parameter that depends on the relative orientation between donor and acceptor dipole and $n$ represents the refractive index of the medium separating donor and acceptor. The parameter $J$ is commonly referred to as spectral overlap integral as it represents the spectral matching of the donor emission and acceptor absorption spectra and is calculated as follows in eq.3[16]:

$$J = \int_0^\infty \overline{f_D}(\lambda)\,\alpha_{M,A}(\lambda)\,\lambda^4\,d\lambda$$

The overlap integral contains the normalized emission spectrum of the donor $\overline{f_D}(\lambda)$ and the molar absorption coefficient of the acceptor $\alpha_{M,A}(\lambda)$, integrated over the wavelength $\lambda$ (grey area in fig.2a). We can use the far-field absorption coefficient of silicon for the near-field (Förster) coupling, because FRET has been measured to also be phonon assisted[7].

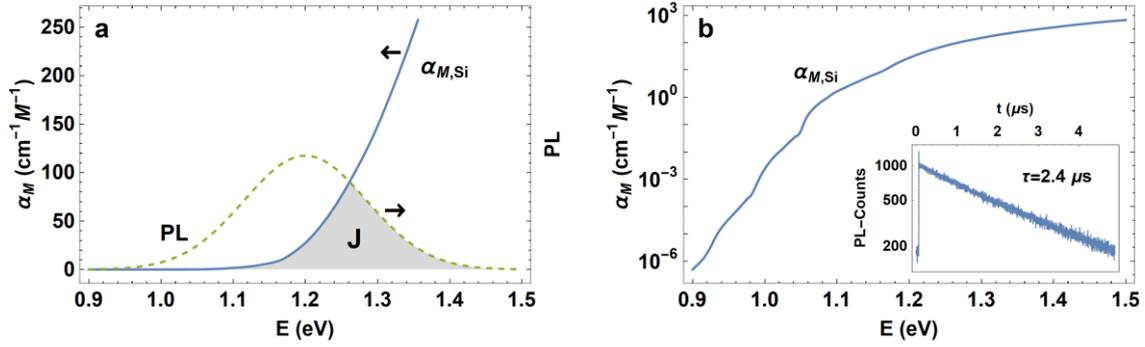

*Figure 2:* **(a)** $\alpha_{M,Si}$ *and PL of 1.2 eV PbS QDs as a function of photon energy. The gray shaded area indicates the spectral overlap between the QD donor and the Si acceptor (J). $\alpha_{M,Si}$ was taken from Green et.al.[29] and the PL spectrum was modelled as a Gaussian centred at 1.2 eV with a FWHM of 200 meV which corresponds to a broadening of $\sigma = 84\ meV$. The PL is scaled by a factor of 25 for visibility.* **(b)** *Molar absorption coefficient of silicon $\alpha_{M,Si}$ as a function of photon energy. The inset shows the measured transient PL lifetime for 1.2 eV PbS QDs.*

Fig.2a depicts $\alpha_{M,Si}$ and $\overline{f_{QD}}$ as a function of energy. The FWHM assumed for the QD PL is $200\ meV$, in agreement with literature[18–20]. The refractive index of the separating medium depends on how one accounts for the contributions of the dielectric functions of the QD itself, the surrounding ligand and the spacer material. Following Yeltik et.al.[7] we consider the average of refractive indices in a straight line from QD to the silicon surface. We approximate the refractive index as constant for different spacer thicknesses. As such, $n_{SiO_2} = 1.45$ is used, which is the index of the SiO₂ spacer layer in between the QDs and the Si bulk. In fact, the QDs and the ligands will also influence the overall refractive index, as the light will be influenced by an effective medium given that the wavelength of emission is much larger than the distances involved in our system. The refractive index of the QDs is well above 1.45, and the refractive index of the organic ligands is between 1.45 for oleic acid (OA)[21] and 1.5 (3-mercatopropyonic acid)[22] for most organic ligands. Inorganic ligands like ZnI₂ are very short so we can neglect their influence on the electromagnetic field. However, since the ligands do not fill the entire volume[19], we deem the approximation of $n = 1.45$ valid for distances larger than $1\ nm$. The orientation parameter $\kappa^2$ depends on the relative transition dipole orientation of donor and acceptor[16]. Since the QDs have rotational symmetry, the dipole orientation in the QDs will be isotropic which yields $\kappa_{iso}^2 = \frac{2}{3}$ [16]. The quantum yield of PbS-QDs depends on various factors including size[23], excitation wavelength[24], QD concentration[24], ligands[25], and whether they are in solution or in solid state. The choice of QD size is important because the corresponding bandgap has to be lower than the tetracene triplet exciton energy and higher than the Si bandgap, to ensure that both transfer processes are downhill in energy. We choose QDs with emission centered at $1.2\ eV$, which corresponds to an average size of $3.4\ nm$[23]. The PLQY for these QDs ranges from 20% – 55%[24] in solution, and up to 15% in films[26]. We determined the radiative lifetime of our 1.2 eV PbS QDs (see experimental methods for details on QD synthesis and PL lifetime measurement) in in octane as $\tau_{PbS} = 2.4\ \mu s$ (inset fig.2b), which is in good agreement with literature[13,18,27]. For a more accurate description of the FRET rate the measured lifetime of the QDs in solution should be replaced by the QD lifetime measured after deposition on quartz, to obtain the reference value for "infinite" donor-acceptor separation $\tau_{D,0}$. We exclude the effects of parasitic absorption in the QD layer because we assume a (sub-) monolayer QD coverage.

In the following we calculate $R_0$ which is the distance for that the transfer efficiency reaches 50% in the dipole-dipole model. While this is not exactly the case for the plane and bulk geometries we will introduce later, $R_0$ is still a useful quantity to estimate separation distances. As can be seen in the upper plot of fig.3, the values for $R_0$ vary from 0.9 $nm$ to up to 1.5 $nm$, depending on the QY and bandgap of the QDs. The steep loss of transfer efficiency below the bandgap of silicon (around 1.12 eV) can be attributed to the exponential decrease in the absorption coefficient. The largest QD bandgap for which energy transfer from triplet excitons in tetracene was observed is 1.23 eV[6], and we indicate the QD bandgap range by the grey area in fig.3. The bottom panel of the same figure shows the FRET efficiency, which obeys a relatively steep slope around 1.2 $eV$, compared to higher bandgaps, suggesting the importance of a careful choice of the QD bandgap. The bottom plot of fig.3 shows FRET efficiencies for 1 $nm$ and 2 $nm$ separation distances, with varying QY. Changes in distance by only 1 $nm$ around $R_0$ lead to an efficiency increase of up to 75 %. The efficiencies at 1 $nm$ separation saturate for bandgaps slightly higher than required in the given geometry at values close to 100 %. It is worth noting that high FRET efficiencies (>65 %) can be achieved at realistic distances (1 $nm$) even for a low QY (20 %).

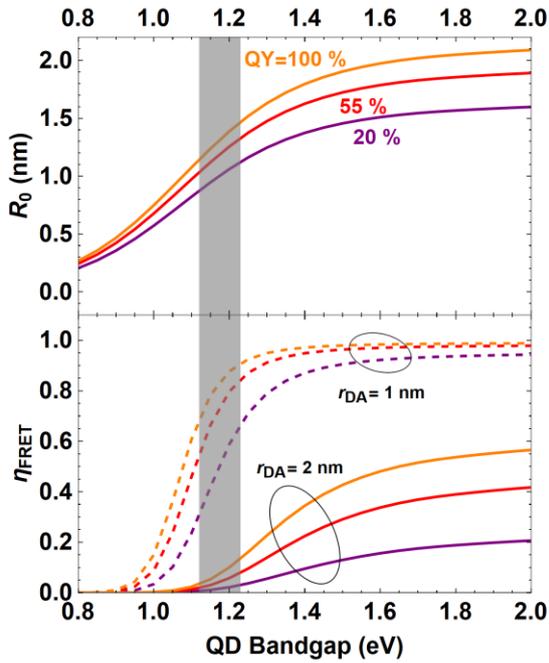

Figure 3: The upper graph shows the QD bandgap dependence of the Förster distance $R_0$ for different quantum yields. In the bottom figure the FRET efficiency as function of QD bandgap is depicted. Dashed lines represent a donor-acceptor distance of 1 nm, solid lines correspond to 2 nm separation. The colours correspond to the same QYs as in the upper figure. The grey shaded region in both plots indicates the bandgap range from 1.12 eV to 1.23 eV, which is the range relevant for the transfer from tetracene into Si.

### Influence of Geometry

Up until now, we have calculated the FRET efficiencies according to a dipole-dipole model that does not take into account the extended nature of the silicon acceptor geometry. We introduce two potentially more accurate descriptions of the FRET rate in our system, in the following referred to as "dipole - infinite plane model" and "dipole – bulk model", similar to earlier approaches[11,28]. The silicon acceptor occupies one half-space instead of being a point-dipole, leading to a modification of eq.2. For the dipole-infinite plane model, the 0-dimensional dipole acceptor is substituted with a 2D acceptor extended over the x-y plane, assuming that the acceptor dipole of FRET mainly resides on the surface of Si (see eq.5).

$$k_{FRET} = \frac{R_0^6}{\tau_{D,0}} \iint_{0,0}^{\infty,2\pi} \frac{x}{\left(R_{DA}(r_{DA},x)\right)^6} \, dx \, d\phi$$

$$= \frac{R_0^6}{\tau_{D,0}} \iint_{0,0}^{\infty,2\pi} \frac{x}{\left(\sqrt{r_{DA}^2 + x^2}\right)^6} \, dx \, d\phi$$

$$= \frac{R_0^6}{\tau_{D,0}} * \frac{\pi}{2r_{DA}^4}$$

Here, $R_{DA}(r_{DA}, x)$ is the distance from the donor dipole to an infinitesimal acceptor dipole. After integration over the Si surface, the rate only depends on the distance-component perpendicular to the surface, thus on $r_{DA}$. The parameterizations used are illustrated in fig.5b.

While this model is closer to the physical reality, it only considers the Si surface. In order to include the Si bulk, we can simply integrate

eq.5 over the half space occupied by Si which leads to eq.6:

$$k_{FRET} = \frac{\pi R_0^6}{2\tau_{D,0}} \int_0^{-\infty} \frac{1}{\left(z'(z,r_{DA})\right)^4} dz$$
$$= \frac{\pi R_0^6}{2\tau_{D,0}} \int_0^{-\infty} \frac{1}{\left(z\left(\frac{n_{Si}}{n}\right) + r_{DA}\right)^4} dz$$
$$= \frac{\pi R_0^6}{6\tau_{D,0}} \left(\frac{n}{n_{Si}}\right) \frac{1}{r_{DA}^3}$$

For the integration $z'(z, r_{DA})$ is split into the integration variable for the half space $z$ and the distance from the donor to the surface of the bulk acceptor $r_{DA}$. The additional prefactor $\frac{n}{n_{Si}}$ arises because we have to consider the refractive index of the part of bulk silicon between the infinitesimal acceptor and the QD donor as part of the separating medium. We use a refractive index of 3.55 for silicon $n_{Si}$, corresponding to the relevant energy region (1.2 $eV$)[29]. For a derivation see the appendix. We note that the prefactor is independent of distance between donor and acceptor. Mathematically this is due to the choice of integration limits and leads to an effective Förster distance $R_{0,eff} = \left(\frac{n_{SiO_2}}{n_{Si}}\right)^{\frac{1}{6}} R_0$.

Fig.4 shows the FRET efficiencies for both models introduced above. From comparison with the bottom panel of fig.3 it becomes obvious that for 2 $nm$ separation the FRET efficiencies are improved considerably up to around 50 % for the dipole – infinite plane model in the relevant region compared to 15 % for the dipole – dipole model, while the values for 1 $nm$ do not change significantly. This occurs due to the different distance dependencies in different models as shown in fig.5. In contrast to the improvement from the 6th to 4th power distance dependence, the change from 4th to 3rd power leads to lower efficiency values for donor-acceptor distances below around 4 $nm$. This observation can also be made by comparing the top and bottom panel of fig.4.

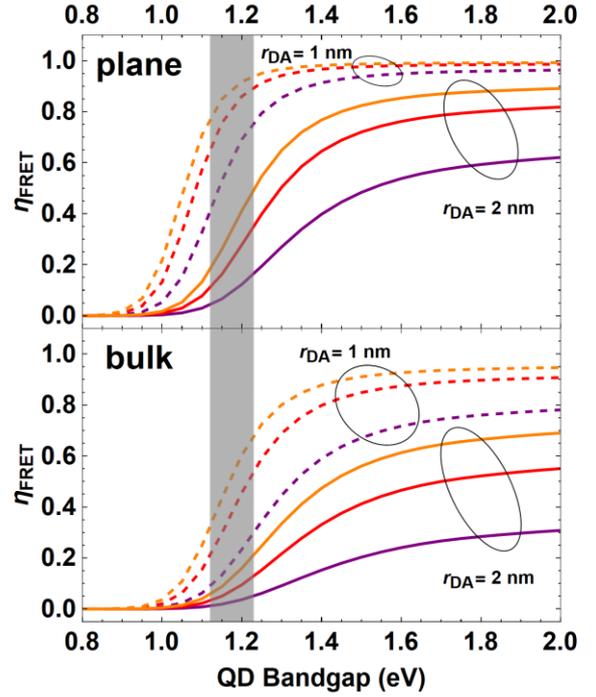

*Figure 4: FRET efficiencies for the "dipole – infinite plane model" (top) and the "dipole – bulk model" (bottom). Dashed and continuous lines represent 1 nm and 2 nm separation, respectively. The grey shaded region indicates the bandgap range of interest. The colours correspond to the same QY values as in fig.3.*

Each of the three models exhibits the highest FRET efficiency in a certain region of distances. Fig.5 illustrates the improved FRET efficiency over most of the range shown in the case of the "dipole – infinite plane model". However, at small distances the point dipole – point dipole model shows the highest efficiencies and the "dipole – bulk model" takes over at distances beyond 5 $nm$, which is not shown in the figure. It is worth noting that the regions defined here have the same limits for any value of $R_0$ which means that in the case of larger Förster distances the bulk-model would be the dominant one. This arises because the difference between the models at short distances $r_{DA}$ becomes marginal for large $R_0$ and with increasing donor-acceptor separation the bulk-model FRET efficiency decreases the slowest.

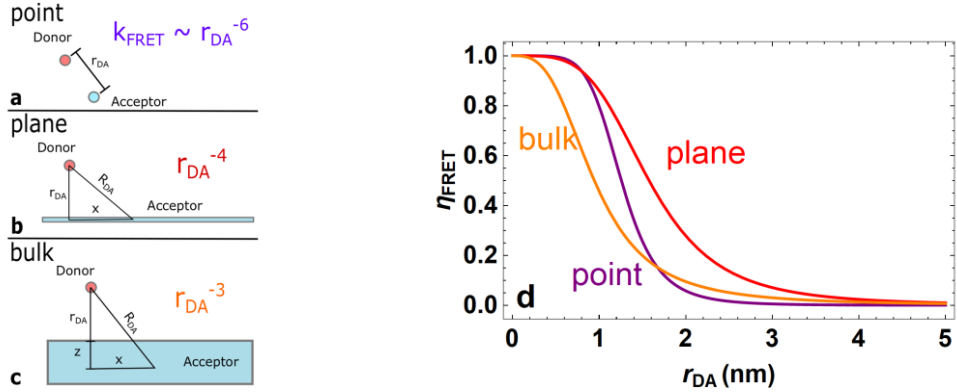

*Figure 5: The three pictures on the left show the three different models ( (a) dipole-dipole, (b) dipole-infinite plane, (c) dipole-bulk) and the corresponding donor acceptor distance dependencies obtained by starting from eq.2 and integrating over a surface or space. The colours indicate which lines in (d) the dependencies correspond to. (d) The graph shows the FRET efficiency for those 3 models at distances in the order of $R_0$. The QD bandgap is 1.2 eV and the QD QY is 55 %, corresponding to $R_0 = 1.26\ nm$.*

At small distances, FRET always out-competes other decay channels. For distances up to about 0.8 nm the point-dipole model yields marginally higher efficiencies as in the plane dipole model the fraction of dipoles at larger lateral distance outweigh the gain by the weaker D-A distance dependence. For distances between $0.8\ nm$ and $4\ nm$ the infinite plane model yields the highest efficiency, while the bulk model only yields higher efficiency when the distance is very large (and overall efficiency low). This is again due to the fact that the dipoles in the bulk are on average further away from the donor, which is only compensated for at larger distances.

With increasing distance, first dipole-plane and then dipole-bulk interactions become relatively stronger as they take into account more area/volume. Which model most accurately describes the distance dependence in our QD-silicon geometry? While the bulk-model represents the geometry more accurately, one could argue that due to the strong distance dependence of FRET the majority of the interaction occurs already at the surface, so the plane-model might be valid after all. However, the spatial extend of the Bloch waves in silicon will ultimately govern the transition geometry.

We note that the mathematical treatment shown here does not take into account that part of the electromagnetic field is reflected by silicon, which leads to a reduced donor lifetime for small distances according to CPS theory[30]. Furthermore, the exciton in the quantum dot could be more accurately described as an extended dipole. The point dipole approximation is no longer valid if the distance between donor and acceptor is on the order of the exciton (QD) size. If the separation between electron and hole ($1.8\ nm$ for PbS QDs[24]) is taken into account, the near field will no longer be accurately described by the $r^{-3}$ dependence used in the FRET derivation. The final step would be the addition of a quantitative description of Dexter transfer[31], which is a possibly competing charge mediated energy transfer. Dexter transfer has an exponential distance dependence, which leads to transfer distances of around $1\ nm$ but it does not depend on the absolute molar absorption coefficient of silicon (only on the spectral shape) which could make Dexter rates comparable with FRET rates in this case.

A factor that greatly affects $k_{FRET}$ is the overlap between QD emission and Si absorption spectra. The QD absorption energy must be lower than the tetracene triplet exciton energy and the emitted energy of the QD must be above the Si bandgap. The broadening of the QD emission spectrum leads to additional losses when the emission spectrum broadens beyond the given limits. Sharper QD emission spectra could be achieved with a QD ensemble with sharper size distribution[32]. Apart from that, the Stokes shift might influence the choice of QD size strongly. We now assumed emission at $1.2\ eV$, which means that the absorption of the QDs would occur at a higher energy. However, the absorption is limited by the fact that tetracene triplet states impose an upper boundary for the QD bandgap of around $1.25\ eV$.

## Conclusion

In conclusion, we showed that FRET from PbS QDs to silicon is possible with sufficiently high FRET efficiencies, even for QDs that have a bandgap close to silicon, and low photoluminescence quantum yield. While efficient FRET is only possible over small separation distances in the order of a few nm, those distances are physically feasible, given careful engineering of the interface.

It is of great importance that the emission and absorption peak of the QDs are between the tetracene triplet exciton energy and the bandgap of Si, with a narrow emission spectrum. Hence, to obtain high FRET efficiency for using singlet fission to improve silicon solar cells, a narrow size distribution of adequate QDs leading to a narrow PL peak and to fine tuning of the bandgap and emission yield of the QDs is necessary. Additionally, the silicon surface needs to be passivated electrically and against oxidation with a very thin (sub-nm) layer. Such layers can be achieved with thin metal oxides[33], or self-assembled monolayers of organic molecules[34]. In case of the organic molecules, they could also act as covalent linkers and passivating ligands for the QDs.

## Experimental Methods

**QD synthesis and passivation**

The colloidal PbS QDs were synthesized by Ruirt Bosma via the hot injection method[35]. In order to obtain the 1.2 eV QDs we measured, the following recipe was followed[36]:

Most chemicals were purchased from Sigma-Aldrich. For those that were not, the distributor will be indicated.

The octadecene is degassed heating to $80\ °C$. A $20\ mL$ syringe is filled with $0.213\ mL$ of bis(trimethylsilyl)sulphide (synthesis grade) together with $10\ mL$ of octadecene (technical grade 90 %) in a glove box ( <$0.5\ ppm\ H_2O$, <$0.5\ ppm\ O_2$) environment. $0.45\ g$ of PbO (99.999 %, Alpha Aesar), $1.34\ g$ of oleic acid (technical grade 90 %) and $14.2\ g$ of octadecene are mixed together in a three-necked Schlenk flask. At a temperature of $95\ °C$ and under vacuum this forms a clear solution. Then, the temperature is increased to around $170\ °C$ in a nitrogen environment. Now, the Schlenk flask containing the lead precursor is transferred to a heating mantle which is at room temperature. As soon as the temperature has reached the injection temperature of $150\ °C$ (for $1.2\ eV$ QDs), the sulphur precursor is injected into the flask with the solution being vigorously stirred. When the solution has cooled down to $35\ °C$, $20\ mL$ of acetone are added.

For surface passivation with $I_2$ we follow Lan et. al.[37]. After the completed synthesis, the QDs are precipitated with acetone in a glovebox. After centrifuging for $4-10\ mins$ at $4000-5000\ rpm$ the residual liquid is disposed of, which is followed by vacuum-drying of the precipitate overnight. The quantum dots are then re-dispersed in toluene ($\geq 99.9\%$) to obtain a concentration of $150\frac{mg}{ml}$. Now a $25\ mM$ iodine (99.999%) in toluene solution is added to the QD solution at a 1:5 ratio and stirred for 24 hours. Afterwards the QDs are precipitated with methanol and centrifuged at $1500-5000\ rpm$ for $2-5\ min$. The residual fluid was disposed of and after a night of vacuum-drying the QDs were dispersed in octane to obtain a $37.5\frac{mg}{ml}$ solution.

Eventually, this solution was diluted with octane to obtain a $4.4\frac{mg}{ml}$ solution, which was used in the lifetime measurements.

**PL lifetime measurement**

The photoluminescence decay of the 1.2eV bandgap PbS QD was measured on a home-built time-correlated single photon counting (TCSPC) system consisting of a $640\ nm$ pulsed laser (PicoQuant LDH-D-C-640) with a repetition rate of $0.2\ MHz$ as an excitation source controlled by a PicoQuant PDL 828 "Sepia II'. The signal was collected by a single-photon avalanche diode (SPAD) detector (Micro Photon Devices, MPD-5CTD) connected to a PicoQuant HydraHarp 400 multichannel picosecond event timer. The laser has a power of $14.6 \, \mu W$ at the used repetition rate. The laser light was filtered out of the collection path by a Chroma ZET 642nf notch filter and a Chroma ET 655lp long pass filter. The TCSPC decays were collected for $5 \, mins$.

## Appendix

**Introduction of bulk silicon as additional separating medium in the dipole – bulk model:**

The distance between the QD donor and the infinitesimal dipole acceptor located at an arbitrary spot somewhere in the silicon bulk can be described as $r_{DA} + z = z'$. Here, $z'$ is the total separation distance and $r_{DA}$ and $z$ are the parts in the SiO₂ medium and in silicon, respectively. For simplicity we now calculate the case for $z'$ perpendicular to the silicon surface (fig.A1), but the following derivation holds for any angle between the donor acceptor connection line and the silicon surface.

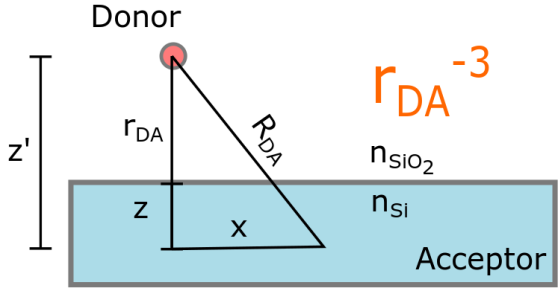

*Fig. A1: Illustration of the geometry for the bulk integration of $k_{FRET}$.*

The total refractive index $n_{tot}$ can be calculated from the effective medium approximation, where $n_{tot}$ is the weighted sum of the two individual indices, for SiO₂ $n_{SiO_2}$ and silicon $n_{Si}$.

$$n_{tot}(r_{DA} + z) = n_{SiO_2} * r_{DA} + n_{Si} * z$$

Solving for $n_{tot}$ leads to:

$$n_{tot} = \frac{n_{SiO_2} * r_{DA} + n_{Si} * z}{r_{DA} + z}$$

$$= n_{SiO_2} \left( \frac{r_{DA} + \frac{n_{Si}}{n_{SiO_2}} z}{r_{DA} + z} \right)$$

*A1*

The obtained expression has to be substituted into a new Förster distance, $R_0'$, following eq.3:

$$R_0'^6 = \frac{9000 \ln(10)}{128 \, \pi^5 \, N_{AV}} * \frac{Q_D \kappa^2 J}{n_{tot}^4}$$

$$= \frac{9000 \ln(10)}{128 \, \pi^5 \, N_{AV}} * \frac{Q_D \kappa^2 J}{n_{SiO_2}^4}$$

$$* \left( \frac{r_{DA} + \frac{n_{Si}}{n_{SiO_2}} z}{r_{DA} + z} \right)^{-4}$$

$$= R_0^6 * \left( \frac{r_{DA} + z}{r_{DA} + \frac{n_{Si}}{n_{SiO_2}} z} \right)^4$$

*A2*

Where $R_0$ is the ordinary Förster distance for SiO₂ as separating medium. This can now be substituted into the equation for the FRET rate which we obtained after integration over the surface:

$$k_{FRET} = \frac{\pi}{2} \int_0^\infty \frac{R_0'^6}{(r_{DA} + z)^4} dz$$

$$= \frac{\pi}{2} R_0^6 \int_0^\infty \frac{1}{(r_{DA} + z)^4} \left( \frac{r_{DA} + z}{r_{DA} + \frac{n_{Si}}{n_{SiO_2}} z} \right)^4 dz$$

$$= \frac{\pi}{2} R_0^6 \int_0^\infty \frac{1}{\left( r_{DA} + \frac{n_{Si}}{n_{SiO_2}} z \right)^4} dz$$

$$= \frac{\pi}{2} R_0^6 \frac{n_{SiO_2}}{n_{Si}} \int_0^\infty \frac{1}{(u)^4} du$$

$$= -\frac{\pi}{6} R_0^6 \frac{n_{SiO_2}}{n_{Si}} \left( 0 - \frac{1}{r_{DA}^3} \right)$$

$$= \frac{\pi}{6} R_0^6 \left( \frac{n_{SiO_2}}{n_{Si}} \right) \frac{1}{r_{DA}^3}$$

*A3*

The equations above show the derivation of the $\frac{n}{n_{Si}}$ prefactor in eq.6 of the main text. For the integration substitution of variables was used with $u = r_{DA} + \frac{n_{Si}}{n} z$.

Acknowledgements

This work is part of the research program of The Netherlands Organization for Scientific Research (NWO). The authors want to thank Ruirt Bosma for the preparation of the PbS QDs and Sven Askes for valuable discussions.